\documentclass[epj]{svjour}
\usepackage{latexsym}
\usepackage{graphics}

\usepackage{bm}
\usepackage{pstricks}

\begin{document}
\input epsf.sty
\title{Monte-Carlo sampling of energy-constrained quantum superpositions in high-dimensional Hilbert spaces}
\author{Frank Hantschel \thanks{\emph{Email:} F.Hantschel@thphys.uni-heidelberg.de} \and Boris V. Fine
\thanks{\emph{Email:} B.Fine@thphys.uni-heidelberg.de}}
\institute{ Institute for Theoretical Physics, University of Heidelberg, Philosophenweg 19, 69120 Heidelberg, Germany }
\date{26 November, 2010}
\abstract{
Recent studies into the properties of quantum statistical ensembles in high-dimensional Hilbert spaces have encountered difficulties associated with the Monte-Carlo sampling of quantum superpositions constrained by the energy expectation value. A straightforward Monte-Carlo routine would enclose the energy constrained manifold within a larger manifold, which is easy to sample, for example, a hypercube. The efficiency of such a sampling routine decreases exponentially with the increase of the dimension of the Hilbert space, because the volume of the enclosing manifold becomes exponentially larger than the volume of the manifold of interest.  The present paper explores the ways to optimise  the above routine by varying the shapes of the manifolds enclosing the energy-constrained manifold.  The resulting improvement in the sampling efficiency is about a factor of five for a 14-dimensional Hilbert space. The advantage of the above algorithm is that it does not compromise on the rigorous statistical nature of the sampling outcome and hence can be used to test other more sophisticated Monte-Carlo routines. The present attempts to optimise the enclosing manifolds also bring insights into the geometrical properties of the energy-constrained manifold itself.
}
\authorrunning{F. Hantschel and B. V. Fine}
\titlerunning{Monte-Carlo sampling of energy-constrained quantum superpositions ...}
\maketitle

\section{Introduction}

Experimental efforts to create quantum computers come increasingly close to controllable manipulations of completely isolated quantum systems consisting of the number of $q$-bits of the order of 10. Although not  macroscopic, such $q$-bit systems have sufficiently large Hilbert spaces, where it becomes increasingly difficult to generate predetermined quantum superpositions. Instead, the experiments are likely to deal with the ensembles of quantum superpositions produced either on purpose or because of experimental constraints. On the theoretical side, the statistical properties of the superpositions of quantum states in many-dimensional Hilbert spaces with a constraint on the energy value (or the expectation value of some other observable quantity) are also of interest for the foundations of quantum statistical physics. At issue here is the applicability of the Boltzmann-Gibbs statistics to completely isolated quantum systems.

Recently, the authors have investigated\cite{boris,borisfrank} the properties of the so-called ``quantum micro-canonical'' (QMC) ensemble\cite{brodypaper} of wave functions having fixed energy expectation value (see also Refs.\cite{wootters,bendermicro,fresch0,fresch1,fresch2,eisertpaper}). For a Hilbert space of dimension $N$ with energy spectrum $\{ E_1, E_2, ..., E_N\}$, the QMC ensemble is formally defined to include all possible wave functions
\begin{equation}
\psi=\sum_{i=1}^N {c_i\phi_i},
\label{psi}
\end{equation}
such that $\sum_{i=1}^N |c_i|^2 E_i = E_{\hbox{\scriptsize av}} $. Here $\phi_i$ are the eigenstates of the system, $c_i$ are the corresponding complex amplitudes, and $E_{\hbox{\scriptsize av}}$ is the energy expectation value. ``All possible wave functions'' in the above definition means that the joint probability distribution of a set of normalized values $\{ c_i \}$ is uniform on the manifold in the Hilbert space constrained by the value of $E_{\hbox{\scriptsize av}}$. 

The primary objective of this paper is to explore and improve the numerical algorithms for sampling the QMC ensemble in high-dimensional Hilbert spaces.

\subsection{Analytical results for the QMC ensemble}
\label{analytical}

The QMC ensemble is different from the conventional micro-canonical ensemble: the latter limits the participating eigenstates to a small energy window around $E_{\hbox{\scriptsize av}}$, while the former does not [see the discussion in Ref.\cite{borisfrank}]. According to the available analytical results\cite{boris,wootters,fresch0}, the QMC ensemble does not lead to  conventional Boltzmann-Gibbs statistics for either the entire isolated quantum system, or for a small subsystem of it.  

For the entire quantum system with sufficiently large but finite Hilbert space of dimension $N$, the QMC ensemble typically leads to the following average occupations of eigenstates\cite{boris,wootters,fresch0}: 
\begin{equation}
\langle p_i \rangle = {1 \over N [1 + \lambda (E_i - E_{\hbox{\scriptsize av}})]},
\label{pav2}
\end{equation}
where $p_i \equiv |c_i|^2$, and $\lambda$ is a parameter that can be obtained numerically for a given energy spectrum and a given value of $E_{\hbox{\scriptsize av}}$.

One can immediately see that, as $E_i$ grows, the decrease in the average occupations of the eigenstates is much slower than exponential. (Were it exponential, such a result would represent the canonical ensemble and lead to the conventional Boltzmann-Gibbs statistics for small subsystems of macroscopic systems.) 

Another remarkable feature of formula (\ref{pav2}) is  that, as a function of $E_i$, it has a pole at $E_{\lambda} = E_{\hbox{\scriptsize av}} - 1/\lambda$. As shown in Ref.\cite{boris},  if the QMC ensemble is considered for a macroscopic system with the energy expectation value $E_{\hbox{\scriptsize av}}$ equal to the average energy of the Boltzmann-Gibbs distribution for the same system at any experimentally realisable temperature, then the above pole approaches extremely closely to the energy of the ground state. As a result, the ground state acquires a macroscopically large occupation, which, if the QMC ensemble were realisable, would signify a new type of condensation phenomenon independent of the statistics of the constitutent particles of the system. For quantum systems that have a large number of states but not too large a number of constitutent particles, the above condensation is not sharp, but rather has a character of smooth crossover as a function of $E_{\hbox{\scriptsize av}}$. 

For a small subsystem of a macroscopic system, the QMC-based result also looks unfamiliar\cite{boris,fresch2}: all states of the subsystem except for the lowest one have the same occupations, while the lowest state has a higher occupation. In other words, the subsystem appears to be in a weighed mixture of two conventional thermal states---the zero temperature state and the infinite temperature state.  

We also note that Eq.(\ref{pav2}) already gives a good description of the QMC ensemble for the values of $N$ of the order of ten\cite{boris,borisfrank}.
In this case, however, the deviations from Eq.(\ref{pav2}) are still noticeable for the occupations of the lowest (or highest) energy levels. However, these deviations can be accounted for with the help of the finite-$N$ corrections introduced in Ref.\cite{borisfrank}. 

Although the QMC-based results appear to contradict everyday experience, this does not mean that the QMC ensemble should be labelled as ``unphysical''.  In everyday experimental situations, one does not deal with completely isolated large quantum systems. The physical significance of the QMC ensemble is that it might be realisable under strong perturbations in isolated quantum systems with small numbers of particles but large numbers of quantum levels, for example, systems of ten $q$-bits.  It is also not clear at present, what are the physical reasons behind the fact that the QMC ensemble is not realised in naturally occurring macroscopic systems. A radical possible explanation of this fact is based on the notion of quantum collapse. 

\subsection{Previous numerical investigations of the QMC ensemble}
\label{previous}

The above mentioned analytical results for the QMC ensemble are based, in part,  on approximations that need to be tested numerically by Monte-Carlo sampling. 

\begin{figure}[t]
\begin{pspicture}(0,0)(3,4)
 \uput[0]{0}(0,2){ \epsfxsize= 2.5in \epsfbox{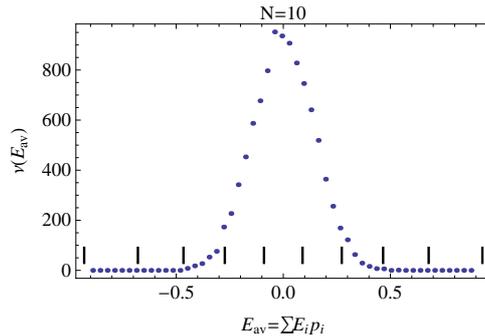} }
\end{pspicture}
\caption{Unnormalised distribution of the values of $E_{\hbox{\scriptsize av}}$ obtained by direct sampling of a 10-dimensional Hilbert space without the energy constraint. The energy levels $E_i$ of the system are indicated above the x-axis. The maximum of the distribution is located at $E_{\hbox{\scriptsize av}} = {1 \over N} \sum_i E_i = 0$. } 
\label{fig:EiPi}
\end{figure}

A straightforward Monte-Carlo routine involves a sampling of the entire Hilbert space with the subsequent selection of the superpositions having the energy expectation values close to $E_{\hbox{\scriptsize av}}$. This routine, however, is rather inefficient.  It is known from analytical calculations\cite{boris,bendermicro} that the occurrence of superpositions with a given energy expectation value $E_{\hbox{\scriptsize av}}$ rapidly decreases as  $E_{\hbox{\scriptsize av}}$ deviates from the arithmetic average of all eigenenergies.  As illustrated in Fig.~\ref{fig:EiPi}, even in modestly large Hilbert spaces there is a range of values of $E_{\hbox{\scriptsize av}}$ that would never appear in the course of such a routine implemented with realistic computational resources.
The larger is the Hilbert space dimension, the smaller is the fraction of all possible values of $E_{\hbox{\scriptsize av}}$ accessible with such a routine.

In our previous work \cite{boris,borisfrank}, in order to examine the statistics for any value of $E_{\hbox{\scriptsize av}}$, we proceeded as follows.

We note that the phases of $c_i$ are not constrained, and hence have uniform probability distribution in the interval $[0, 2 \pi )$. At the same time, the variables $|c_i|$ can be substituted by the eigenstate occupation variables 
$p_i = |c_i|^2$. It can be shown\cite{boris} that the uniform joint probability distribution for the normalised set of $\{ c_i \}$ translates into a uniform distribution on a manifold $M$ defined  in the space of variables $\{p_i\}$ by the following conditions:
\begin{eqnarray}
\sum_{i=1}^{N}  E_i p_i &= E_{\hbox{\scriptsize av}} \label{energCond} \\
\sum_i^N p_i &= 1 \label{normCond} \\
p_i &\geq 0,\ \ \forall i \label{posCond}.
\end{eqnarray}

The advantage of working in the space of variables $\{ p_i \}$ is that all constraints (\ref{energCond}-\ref{posCond}) are hyperplanes, and hence the resulting manifold has the character of hyper-polyhedron with flat faces.  For comparison, in the space of variables  $c_i$ the manifold is curved --- consequence of the hypersphere normalisation condition $\sum_i^N |c_i|^2= 1$ . 

In \cite{boris,borisfrank}, in order to examine the statistics corresponding to any given value of $E_{\hbox{\scriptsize av}}$, we performed the Monte Carlo sampling within a rectangular $(N-2)$-dimensional box around the above manifold in the subspace constrained by hyperplanes (\ref{energCond}, \ref{normCond}). This sort of algorithm, while being able to access any value of $E_{\hbox{\scriptsize av}}$, is still relatively inefficient. As the dimension of the Hilbert space increases, the volume of the enclosing box becomes exponentially larger than the volume of the manifold of interest, and hence the acceptance rate for the Monte-Carlo points becomes very small. The purpose of the present work was to explore how much one can improve the acceptance rate by reorienting the above rectangular box with the appropriate resizing of its linear dimensions, or by choosing a non-rectangular parallelogram-like box.

Other algorithms to sample the QMC ensemble have been meanwhile proposed\cite{fresch0,eisertpaper} and, in principle, can sample the QMC ensemble more efficiently. These algorithms guide the sampling on the basis of available analytical results.  However, since some of these analytical results are themselves of approximate nature, it remains to be shown, that such algorithms lead to a fair representative sampling of the QMC ensemble. Given that some of the properties of the QMC ensemble, such as the condensation to the ground state, are rather non-intuitive and have unclear sensitivity to the numerical uncertainties, it is highly desirable to remove from the numerical investigations any doubts about the fair character of the Monte-Carlo sampling routines.  The clear advantage of the algorithms considered in the present paper is that, whenever they produce sufficient statistics, this statistics is guaranteed to represent the true QMC ensemble. Therefore, the relatively slow algorithms described below can be used to test faster algorithms.

We also note that the present effort to optimise the choice of the sampling box around the manifold $M$ reveals interesting insights into the geometry of this manifold itself.

In a broader mathematical context, the problem of computing the volume of a generic convex high-dimensional polyhedron is known to be NP-hard\cite{Dyer-88}. This and related optimisation problems are the subject of active ongoing research---see, for example, Refs.\cite{Lawrence-91,Burger-00,Simonovits-03}. In comparison with the generic polyhedron problem, the case considered in the present paper is somewhat simpler, because, as shown below, we can easily identify vertices, edges and faces of the polyhedron of interest.

\

In the rest of the paper, Section \ref{ch:basicalg} describes the basic idea of the Monte-Carlo sampling algorithm, Section~\ref{ch:vertices} describes the vertices of the manifold $M$, Sections~\ref{ch:1stalg} and \ref{ch:2ndalg} introduce two improved algorithms, Section \ref{ch:performance} presents the performance tests for the algorithms considered, and, finally, Section \ref{ch:summary} summarises the results presented in this paper.

\section{Monte-Carlo algorithms}

\subsection{Basic algorithm}
\label{ch:basicalg}

Our basic algorithm for the Monte-Carlo sampling of manifold $M$ defined by Eqs.(\ref{energCond}-\ref{posCond}) is based on putting a $(N-2)$-dimensional box referred to as  $B$ around manifold $M$. The box $B$ should lie in the the $(N-2)$-dimensional hyperplane $A$ formed by the intersection of the energy and the normalisation hyperplanes given by Eqs.(\ref{energCond}) and (\ref{normCond}), respectively.  Different algorithms discussed in subsections~\ref{ch:1stalg} and \ref{ch:2ndalg} differ only by the shape and the orientation of box $B$. In this subsection we describe all the common elements of these algorithms, which we call the ``basic algorithm". It consists of the following steps:

1) Define the  ``standard'' coordinate system in the space of variables $\{ p_i \}$ with the origin at point $(0,0,...,0)$  and the set of $N$ basis vectors: ${\mathbf e}_1 = (1,0,0,...,0)$, ${\mathbf e}_2 = (0,1,0,...,0)$, ...,   ${\mathbf e}_N = (0,0,0,...,1)$.

2) Define the ``modified'' coordinate system, where the origin is chosen on one of the vertices of manifold $M$ (that is, in the hyperplane A), and the basis vectors are selected as follows: The first two vectors are orthogonal to the hyperplane A and denoted as ${\mathbf b}_{\hbox{\scriptsize nrm}}$ and ${\mathbf b}_E$. Vector ${\mathbf b}_{\hbox{\scriptsize nrm}}$ is chosen perpendicular to the normalisation hyperplane (\ref{normCond}), that is, ${\mathbf b}_{\hbox{\scriptsize nrm}}=(1,1,1,...,1)/\sqrt{N})$.   Vector ${\mathbf b}_E$ is obtained by orthonormalising vector $(E_1, E_2, E_3,..., E_N)$ with respect to the ${\mathbf b}_{\hbox{\scriptsize nrm}}$ with the help of the Gram-Schmidt procedure.  The remaining $N-2$ basis vectors, to be denoted as ${\mathbf b}_1, {\mathbf b}_2, ..., {\mathbf b}_{N-2}$, are all orthogonal to ${\mathbf b}_{\hbox{\scriptsize nrm}}$ and ${\mathbf b}_E$, that is, they  lie in the hyperplane $A$, but they are not necessarily orthogonal to each other. Their specific choice depends on the particular realization of the algorithm as described in the following sections.  

3) Find all vertices of manifold $M$ in the standard coordinate system using formulas given in subsection~\ref{ch:vertices}.

4) Transform the coordinates of all vertices to the modified coordinate system.

5) Define the orientation of box $B$ in the modified coordinate system; choose the linear dimensions of box $B$ that are minimally sufficient to enclose all vertices of manifold $M$ within the box.

6) Randomly sample points within box $B$ in the modified coordinate system.

7) Select the points belonging to manifold $M$. The selection criterion is the following: Each sampled point should be transformed to the standard coordinate system and then accepted only if all standard coordinates are non-negative, meaning that the positivity condition (\ref{posCond}) is satisfied.

\

Denoting the number of sampled points as $n_s$, the number of accepted points as $n_a$, the $(N-2)$-dimensional volume of manifold $M$ as $V_M$ and the volume of the surrounding box as $V_B$, we express the acceptance rate of the algorithm as follows 
\begin{equation}
r = {n_a \over n_s} = { V_M \over V_B}.
\label{eqn:r}
\end{equation}
From our experience, a reasonable acceptance rate for practical computation should be larger than $10^{-7}$. 

Below we consider the following choices for the enclosing box $B$:

1) In subsection~\ref{ch:1stalg}, we choose the box $B$ to be of rectangular shape and optimise its orientation. Changing the box orientation modifies its dimensions and hence, once optimised, can reduce its volume. The smaller the volume of the box, the larger the acceptance rate $r$. We refer to the resulting algorithm as "R-algorithm" (R for rectangle).

2) In subsection \ref{ch:2ndalg} we enclose  the manifold $M$ within a non-rectangular parallelogram-like box and then optimise the choice of this box.  We call the resulting algorithm "NR-algorithm" (NR for non-rectangular).

Both R- and NR-algorithms require finding the  vertices of manifold $M$ --- step 3 of the basic algorithm. This step is described in the next subsection.

\subsection{Vertices of manifold $M$}
\label{ch:vertices}

In the standard coordinate system, each vertex of the manifold $M$ has only two non-zero coordinates\cite{boris}. Moreover, the fact that the $i$-th and the $j$-th coordinates of the vertex are not zeroes uniquely identifies the vertex. Hence the vertices can be labelled by the indices of the non-zero coordinates.  The coordinates of vertex $v_{i,j}$ are
\begin{equation}
\label{eqn:vertices}
v_{i,j}=(0, ..., 0,p_i,0,...,0,p_j,0,...,0),
\end{equation}
where
\psset{unit=1cm}
\begin{eqnarray}
\label{eqn:vertexPi}
p_i &= \frac{E_{\hbox{\scriptsize av}}-E_j}{E_i-E_j} \\
\label{eqn:vertexPj}
p_j &= \frac{E_{\hbox{\scriptsize av}}-E_i}{E_j-E_i}.
\end{eqnarray}
We adopt the convention that the order of indices in $v_{i,j}$ is always such that  $E_i < E_j$.
Since, according to condition (\ref{posCond}), both $p_i$ and $p_j$ should be non-negative, the vertex $v_{i,j}$ exists only when $E_i \leq E_{\hbox{\scriptsize av}}\leq E_j$.
Denoting the numbers of energy levels  below and above $E_{\hbox{\scriptsize av}}$ by $K$ and $L$, respectively, we can thus express the total number of vertices of manifold $M$ as
\begin{equation}
\label{eqn:noVertices}
N_v=K\cdot L.
\end{equation}
Obviously $K+L=N$. Therefore, dependent on the position of $E_{\hbox{\scriptsize av}}$ within the energy spectrum, the value of $N_v$ changes between $N-1$ and $N^2/4$ (for even $N$).

In order to implement the NR-algorithm, we will also need to know which vertices $v_{k,l}$ are connected to a given vertex $v_{i,j}$ by a linear edge of manifold $M$. The criterion here is that either $k = i$ or $l=j$. Therefore,
 the total number $N_e$ of edges originating from each vertex of $M$ is
\begin{equation}
\label{eqn:noNeighbours}
N_e=(K-1)+(L-1)=N-2
\end{equation}
This number is thus exactly equal to the dimension of manifold $M$ for any value of $E_{\hbox{\scriptsize av}}$.

\subsection{Rectangular box: R-algorithm}
\label{ch:1stalg}

For the R-algorithm, we choose box $B$ to be a high-dimensional hyper-rectangle (orthotop) enclosing manifold $M$.  In this algorithm, the modified basis vectors 
$\{ {\mathbf b}_1, {\mathbf b}_2, ..., {\mathbf b}_{N-2} \}$ specified at step 2 of the basic algorithm are chosen to be orthonormal, and then the edges of box $B$ are oriented along these vectors. The basis vectors $\{ {\mathbf b}_1, {\mathbf b}_2, ..., {\mathbf b}_{N-2} \}$ are constructed with the help of the Gram-Schmidt procedure, which requires $N-2$ non-collinear input vectors ${\mathbf g}_1, {\mathbf g}_2, ..., {\mathbf g}_{N-2}$ defined in the standard basis. Each new input vector ${\mathbf g}_i$ is orthogonalized first with respect to ${\mathbf b}_{\hbox{\scriptsize nrm}}$ and ${\mathbf b}_E$, and then with respect to all already available vectors ${\mathbf b}_j$. 

\begin{figure}[t]
\begin{pspicture}(-0.3,0)(14,6)

\psline[linewidth=1.2pt]{-}(0.5,1)(2.5,1)(1.5,5)(0.5,1)
\psline[linewidth=1.2pt,linecolor=red]{-}(0.5,1)(2.5,1)(2.5,5)(0.5,5)(0.5,1)

\psline[linewidth=1.2pt]{-}(4.5,1.5)(6.5,1.5)(5.5,5.5)(4.5,1.5)
\psline[linewidth=1pt,linecolor=red]{-}(5.5,0.5)(3,3)(5.5,5.5)(8,3)(5.5,0.5)

\rput(2.8,4){\textcolor{red}{$B_1$}}
\rput(4,4.5){\textcolor{red}{$B_2$}}
\rput(1.5,2){$M$}
\rput(5.5,2.5){$M$}

\end{pspicture}
\caption{(Colour online) Cartoon representing two different choices of boxes $B_1$ and $B_2$ (red rectangles) enclosing the same manifold $M$ (black triangle). Obviously the volume of box $B_1$ is smaller than the volume of box $B_2$.}
\label{fig:ortho}
\end{figure}
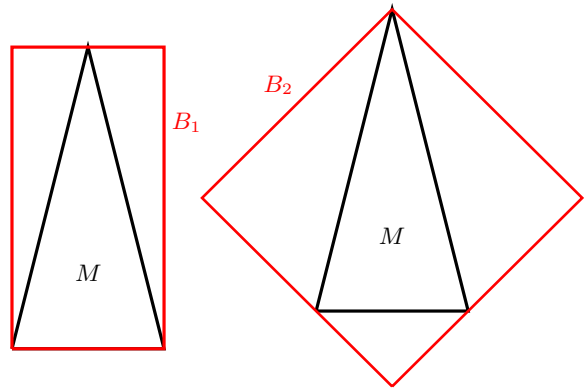

The orientation of the resulting dimensions of box $B$ depends on the choice of the input vectors ${\mathbf g}_i$ and on their sequence in the above Gram-Schmidt procedure. These input vectors can be chosen randomly, but such a choice is a priori unlikely to be optimal:  indeed, it is not (see Section~\ref{ch:performance}). As illustrated in Fig.~\ref{fig:ortho} by a two-dimensional cartoon, different orientations of rectangular boxes around a polygon can clearly lead to different box volumes, and hence, according to formula~(\ref{eqn:r}), different acceptance rates. Such a difference might be more dramatic in higher dimensions. 

We have performed a partial optimisation of the box orientations on the basis of the following idea:  
As one can see in Fig.~\ref{fig:ortho}, a possibly economical way to put a box around the polyhedron $M$ is to pick one face of this polyhedron to coincide with one face of the box. 

To utilise this idea, we note that, according to Eqs.(\ref{energCond}-\ref{posCond}) each $(N-3)$-dimensional face of $M$ is determined by the intersection of the hyperplane $A$ with one of the $(N-1)$-dimensional hyperplanes given by condition $p_i = 0$, that is, each face can be labeled as $F_i$ by the index of the corresponding eigenstate.

Selecting the first input vector for the Gram-Schmidt procedure as ${\mathbf g}_1 = {\mathbf e}_i$ causes ${\mathbf g}_1$ to be orthogonal to $F_i$ thus guaranteeing that one of the resulting box faces will coincide with $F_i$.

Since each face $F_i$ is associated with a different energy level $E_i$, different faces $F_i$ are not equivalent.  Therefore, the
question remains:  Which of the $N$ faces $F_i$ should be chosen to coincide with a face of box $B$?  Or, equivalently: Which of the $N$ natural basis vectors ${\mathbf e}_i$ should be selected as ${\mathbf g}_1$? 
It should be further noted, that different faces $F_i$ are, in general, non-orthogonal to each other. Therefore, once one face $F_i$ coincides with a face of the rectangular box, such a coincidence cannot occur for other faces. Nevertheless, the resulting box volume may depend on the sequence of the remaining vectors ${\mathbf g}_i$ used for the Gram-Schmidt procedure. 

Below we investigated  a limited group of boxes  obtained by assigning each Gram-Schmidt input vector ${\mathbf g}_i$ to be equal to one of the standard basis vectors ${\mathbf e}_j$. 
The investigation was conducted for the following 10-level energy spectrum representing a crudely discretised version of the Gaussian density of states symmetric with respect to zero energy:
\begin{eqnarray}
&\{E_i\}_{i=1..10}=\{-0.929,-0.679,-0.466,-0.273,  
\nonumber
\\ \label{spectrum10}
&-0.09,0.09,0.273,0.466,0.679,0.929\}
\end{eqnarray}
It is the same spectrum as the one shown in Fig.~\ref{fig:EiPi}.

All possible input sequences of type 
\begin{eqnarray}
{\mathbf g}_1 &=& {\mathbf e}_{i_1},
\nonumber
\\
{\mathbf g}_2 &=& {\mathbf e}_{i_2},
\nonumber
\\
&...&,
\label{g1ei}
\\
{\mathbf g}_{N-2} &=& {\mathbf e}_{i_{N-2}},
\nonumber
\end{eqnarray}
 have been tried, and the resulting box volumes for each sequence were obtained.   All input sequences were divided into $N$ groups with  each group having the same standard basis vector ${\mathbf e}_i$ used as the first Gram-Schmidt input vector ${\mathbf g}_1$.  For each group, the minimum volume of box $B$ was found over all possible combinations of vectors ${\mathbf e}_j$ used in the rest of the Gram-Schmidt sequence.  The results corresponding to $E_{\hbox{\scriptsize av}} = -0.3$ are summarised in Table~\ref{tab:Rvolumes}. It was found that the global minimum volume was encountered in nine out of ten groups, the exception being the group corresponding to ${\mathbf g}_1 = {\mathbf e}_1$ with the minimum volume exceeding the global minimum by about 20 per cent. Within each group, there are multiple occurrences of the minimum volume. The number of these occurrences is referred to in  Table~\ref{tab:Rvolumes} as the ``degeneracy factor". 

\begin{table}[tb]
\begin{center}
\begin{tabular}{|c|c|c|} \hline
group index  & min. box & degeneracy \\
$i$ & volume $V_i$  & factor \\
\hline
 1 & 0.0992 & 6  \\
\hline
 2 & 0.0848 & 1872  \\
\hline 
 3 & 0.0848 & 6192 \\
\hline
 4 & 0.0848 & 6192  \\
\hline
 5 & 0.0848 & 7056  \\
\hline
 6 & 0.0848 & 7056  \\
\hline
 7 & 0.0848 & 6912  \\
\hline 
 8 & 0.0848 & 6480 \\
\hline
 9 & 0.0848 & 4320  \\
\hline
 10 & 0.0848 & 72  \\
\hline
\end{tabular}
\end{center}
\caption{Table summarising the investigation of rectangular box volumes for energy spectrum (\ref{spectrum10}) with $E_{\hbox{\scriptsize av}} = -0.3$. 
All Gram-Schmidt input sequences of form~(\ref{g1ei}) have been divided into groups according to the the index of the standard basis vector ${\mathbf e}_i$ used to define the first Gram-Schmidt input vector in ${\mathbf g}_1 = {\mathbf e}_i$. The left column labels these box groups. The middle column gives the value of the minimum box volume $V_i$ for each group.
The right column contains the number of times the minimum volume $V_i$ is encountered within the group.
}
\label{tab:Rvolumes}
\end{table}
\begin{figure}[t]
\begin{pspicture}(0,0)(5,4)
 \uput[0]{0}(0,2){ \epsfxsize= 3in \epsfbox{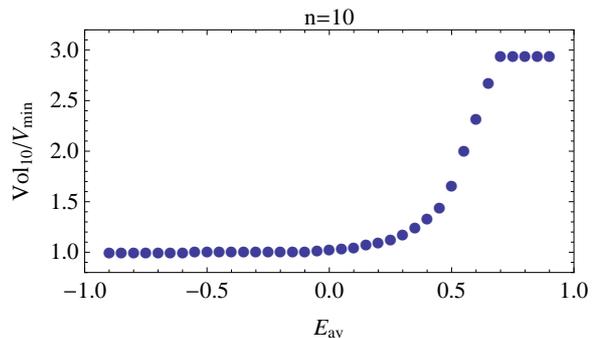} }
\end{pspicture}
\caption{Plot of $V_{10}/V_{\hbox{\scriptsize min}}$ as a function of $E_{\hbox{\scriptsize av}}$, where $V_{10}$ is the minimum volume for the box group characterised by ${\mathbf g}_1 = {\mathbf e}_{10}$, and $V_{\hbox{\scriptsize min}}$ is the minimum among all boxes considered (degeneracy factor).} 
\label{fig:fCvsEav}
\end{figure}

For other values of $E_{\hbox{\scriptsize av}}$, the qualitative character of the results is the same as for $E_{\hbox{\scriptsize av}}=-0.3$. Namely, the choice of ${\mathbf g}_1 = {\mathbf e}_1$ for $E_{\hbox{\scriptsize av}}<0 $ and  ${\mathbf g}_1 = {\mathbf e}_{10}$ for $E_{\hbox{\scriptsize av}}>0 $ does not allow one to reach the global minimum for the volume of box $B$. Figure~\ref{fig:fCvsEav} presents the ratio $V_{10}/V_{\hbox{\scriptsize min}}$ as a function of  $E_{\hbox{\scriptsize av}}$, where $V_{10}$ is the minimum volume  for the group characterised by ${\mathbf g}_1 = {\mathbf e}_{10}$   and $V_{\hbox{\scriptsize min}}$ is the global minimum volume for all groups.  

Another generic feature apparent from Table \ref{tab:Rvolumes} is the greater chance of encountering the global minimum (larger degeneracy factor) for groups with ${\mathbf g}_1 = {\mathbf e}_i$, such that the corresponding energy level $E_i$ is located in the middle of the spectrum. 

Within each group, the influence of the choice of consecutive input vectors ${\mathbf g}_2, ..., {\mathbf g}_{N-2}$ on the resulting box volume is complicated to describe for the present 10-level spectrum, and we do not attempt it here.  We have, however, also conducted similar investigations for  smaller spectra  with $N=5$ and $N=6$, where a clear picture emerged. In the both $N=5$ and $N=6$ cases, we have found that (i)   the last input vector ${\mathbf g}_{N-2}$
does not confine the volume in any way; and (ii)  the  minimal box volume is realised, whenever neither of the vectors ${\mathbf g}_1, ..., {\mathbf g}_{N-3}$  is assigned to be ${\mathbf e}_1$ or ${\mathbf e}_N$.  Therefore, for each  input group where the global minimum volume appears (that is, for  ${\mathbf g}_1 = {\mathbf e}_{i_1}$ with $i_1 = 2,...,N-1$),  the degeneracy factor is  equal to $3 (N-3)!$. While such a  result is valid for $N=5$ and $N=6$,  it does not hold for $N=10$ (see  Table~\ref{tab:Rvolumes}).

Our final prescription  for the R-algorithm is the following:
At step 2 of the basic algorithm, we construct orthonormal modified basis vectors ${\mathbf b}_1,..., {\mathbf b}_{N-2}$ using the Gram-Schmidt orthogonalization procedure with input vectors ${\mathbf g}_1 = {\mathbf e}_{i_1}$,..., ${\mathbf g}_{N-2} = {\mathbf e}_{i_{N-2}}$, where none of the vectors ${\mathbf e}_{i_k}$ are equal to either ${\mathbf e}_1$ or ${\mathbf e}_N$, and otherwise vectors ${\mathbf e}_{i_k}$ appear in the order of the proximity of the corresponding energies $E_{i_k}$ to the arithmetic average of all energies in the spectrum. At step 5, we choose a rectangular box with the edges oriented along vectors ${\mathbf b}_1, ..., {\mathbf b}_{N-2}$. 

In each calculation, the above prescription can be partially controlled by doing random sampling of a large number of Gram-Schmidt input sequences $\{{\mathbf g}_k = {\mathbf e}_{i_k}\}$ and then checking that the volumes of the resulting boxes are not smaller than the volume of the ``prescription box''.  All such tests performed in the specific cases presented in Section~\ref{ch:performance} have supported the above prescription.

\subsection{Non-rectangular box: NR-algorithm}
\label{ch:2ndalg}

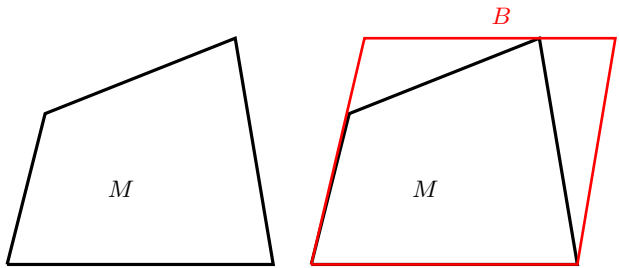
\begin{figure}[t]
\begin{pspicture}(-0.2,0)(7,4)

\psline[linewidth=1.2pt]{-}(0,0.5)(0.5,2.5)(3,3.5)(3.5,0.5)(0,0.5)

\psline[linewidth=1.2pt]{-}(4,0.5)(4.5,2.5)(7,3.5)(7.5,0.5)(4,0.5)
\psline[linewidth=1pt,linecolor=red]{-}(4,0.5)(4.7,3.5)(8,3.5)(7.5,0.5)(4,0.5)

\rput(6.5,3.8){\textcolor{red}{$B$}}
\rput(1.5,1.5){$M$}
\rput(5.5,1.5){$M$}

\end{pspicture}
\caption{(Colour online) Cartoon of parallelogram-like box $B$ (red parallelogram) enclosing the manifold $M$ (black  triangle).
The large dot denotes the ``origin vertex" of $M$, while the arrows indicate the basis vectors used to create $B$.}
\label{fig:nonortho}
\end{figure}

The NR-algorithm uses non-rectangular parallelogram-like box for $B$. In order to define such a box, one needs to specify one of the box vertices, which we refer to as the ``origin vertex", and the $N-2$ non-orthogonal edges originating from this vertex. As indicated by Eq.(\ref{eqn:noVertices}), each vertex of manifold $M$ also has exactly $N-2$ edges originating from it. Therefore, the origin vertex of box $B$ is chosen to coincide with one of the vertices of manifold $M$, and then the corresponding $N-2$ edges of the manifold $M$ determine the edge directions for box $B$. A two-dimensional cartoon of such an arrangement is shown in Fig.~\ref{fig:nonortho}.  The origin vertex also becomes the origin of the modified coordinate system, and the vectors of the modified basis are oriented along the same  $N-2$ edge directions of box $B$. 

Such a general algorithm still leaves the freedom of choosing a vertex of manifold $M$ as the origin vertex of box $B$. The number of vertices of manifold $M$ is given by 
Eq.(\ref{eqn:vertexPj}). Below we explore the dependence of the box volume on the choice of the origin vertex for the energy spectrum (\ref{spectrum10}). A possible algorithm for the calculation of the volume of a parallelogram-like box is given in Appendix~\ref{ch:directcalc}

\begin{figure}[t]
\begin{pspicture}(0,0)(5,5.5)

\put(0.8,-1.2){\epsfxsize= 3.5in \epsfbox{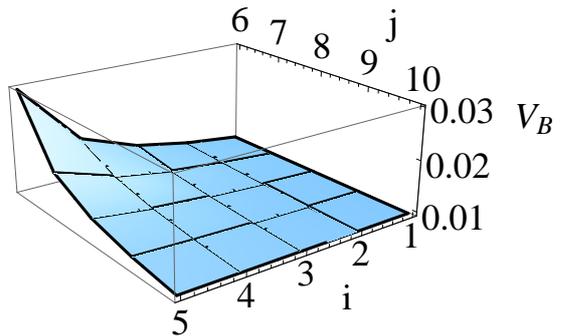} }

\end{pspicture}
\caption{Volumes of parallelogram-like boxes $V_B$  for different choices of the origin vertex $v_{i,j}$ for the Gaussian spectrum of $N=10$ and $E_{\hbox{\scriptsize av}}=0$. Coordinates $i$ and $j$ are the indices of the chosen vertex $v_{i,j}$.
choice.}
\label{fig:2nd_theovol3d}
\end{figure}

As explained in Section~\ref{ch:vertices}, each vertex is labeled as $v_{i,j}$, where the two indices are such that $E_i < E_{\hbox{\scriptsize av}}$ and $E_j > E_{\hbox{\scriptsize av}}$. 
Figure~\ref{fig:2nd_theovol3d} presents the results of box volume calculations for for $E_{\hbox{\scriptsize av}}=0$. In this case, $E_5< E_{\hbox{\scriptsize av}} < E_6$. Therefore, there exist 25 vertices $v_{i,j}$ with index $i$ taking one of the values $\{1,2,3,4,5\}$, while index $j$ may take values $\{6,7,8,9,10\}$. The results are presented as a three-dimensional plot, where two horizontal axes represent vertex indices $i$ and $j$ and the vertical coordinate represents the resulting volume. 
In this figure, the ratio $R_{\mathrm{vol}}$ of the  maximum volume $V_{\mathrm{max}}$ over the minimum
volume $V_{\mathrm{min}}$ is
\begin{equation}
R_{\mathrm{vol}} \equiv \frac{V_{\mathrm{max}}}{V_{\mathrm{min}}}=3.10627
\end{equation}
As one can see from Fig.~\ref{fig:2nd_theovol3d},
there exist nine cases realizing the minimum volume.  They correspond to  vertices of the form $v_{1,j}$ with arbitrary $j$, or  $v_{i,10}$ with arbitrary $i$. The maximum volume corresponds to only one vertex $v_{5,6}$ with two respective levels bracketing $E_{\hbox{\scriptsize av}}$ from below and above. Qualitative arguments explaining, why it is expected that such a situation is special, are given in Appendix \ref{ch:maxvolbox}.
 
We have further investigated the characteristics of box volumes, for several values of $E_{\hbox{\scriptsize av}}$ and collected the results
for $V_{\mathrm{min}}$ and $R_{\mathrm{vol}}$ in Table~\ref{tab:theovol}. This table demonstrates that (i) the ratio $R_{\mathrm{vol}}$ is the same for different values of $E_{\hbox{\scriptsize av}}$ but agreeing values of $K$ and $L$; and (ii) $R_{\mathrm{vol}}$ decreases with decreasing number of vertices (equal to $K L$).

\begin{table}[t]
\begin{center}
\begin{tabular}{|c|c|c|c|c|} \hline
$E_{\hbox{\scriptsize av}}$ & $K$ & $L$ & $V_{\mathrm{min}}$ & $R_{\mathrm{vol}}$ \\
\hline
0 & 5 & 5 & $9.93\,10^{-3}$ & 3.11 \\
-0.04 & 5 & 5 & $9.86\,10^{-3}$ & 3.11 \\
-0.08 & 5 & 5 & $9.64\,10^{-3}$ & 3.11 \\
-0.10 & 4 & 6 & $9.27\,10^{-3}$ & 2.97 \\
-0.46 & 3 & 7 & $8.20\,10^{-3}$ & 2.55 \\
-0.5 & 2 & 8 & $4.60\,10^{-3}$ & 1.88 \\
\hline
\end{tabular}
\end{center}
\caption{Properties of the volumes of parallelogram-like boxes $B$ for the 10-level energy spectrum (\ref{spectrum10}) as a function of $E_{\hbox{\scriptsize av}}$. The parameters of the table are defined in the text.
}
\label{tab:theovol}
\end{table}

Our final prescription for the NR-algorithm is to choose vertex $v_{1,N}$ as the origin vertex. (Since, from the computational viewpoint,  the number of possible choices of the origin vertex is not large,  one can always check the above prescription by computing the box volumes for all possible choices.)

\section{Performance of R- and NR-algorithms}
\label{ch:performance}

This section compares the performance of the algorithms presented in section \ref{ch:1stalg} and \ref{ch:2ndalg} by
looking at their acceptance rates. The following conditions were chosen:

1) Number of states $N$ was equal to 10, 12, or 14 ;

2) Energy spectra represented discretised realisations of the Gaussian density of states similar to spectrum (\ref{spectrum10}) with arithmetic average of all energies equal to zero and root-mean-squared deviation from zero equal to $1/\sqrt{2}$ .

3) The average energy $E_{\hbox{\scriptsize av}}$ was chosen between 0 and $0.8  E_1$. ($E_1$ is the minimum eigenenergy of the spectrum.)

\

Table~\ref{tab:performance} contains the acceptance rates for three different algorithms: (i) non-optimised R-algorithm with random isotropic selection of Gram-Schmidt input vectors ${\mathbf g}_1, ..., {\mathbf g}_{N-2}$;  (ii) optimised R-algorithm with prescription for the input Gram-Schmidt vectors given at the end of subsection \ref{ch:1stalg}; and (iii) optimised NR-algorithm with prescription for choosing the origin vertex given at the end of subsection \ref{ch:2ndalg}.
We use non-optimised R-algorithm as a benchmark for inefficient choice of box $B$, which allows us to judge the effect of optimising box $B$.

\begin{table*}[t]
\begin{center}
\footnotesize\begin{tabular}{|c|c|c|c|c|c|c|} \hline
\label{tab:performance}
 & & &R-algorithm& & \\
 & & &(not optimised)& & \\
$N$ & $E_{\hbox{\scriptsize av}}$ & ${E_{\hbox{\scriptsize av}} \over E_1}$ &random input& R-algorithm &NR-algorithm\\
 & & &to the& (optimised) &(optimised) \\
 & & &Gram-Schmidt pr.& &\\
\hline
 10 & 0 & 0 & $1.3\cdot10^{-4}$ & $1.6\cdot10^{-4}$ & $1.6\cdot10^{-4}$ \\
 12 & 0 & 0 & $2.1\cdot10^{-6}$ & $2.8\cdot10^{-6}$ & $2.7\cdot10^{-6}$ \\
 14 & 0 & 0 & $3\cdot10^{-8}$ & $3\cdot10^{-8}$ & $4\cdot10^{-8}$ \\

\hline

 10 & -0.46 & 0.5 & $1.6\cdot10^{-5}$ & $5\cdot10^{-5}$ & $1.8\cdot10^{-5}$ \\
 12 & -0.63 & 0.5 & $2.2\cdot10^{-7}$ & $6.6\cdot10^{-7}$ & $7.2\cdot10^{-7}$ \\
 14 & -0.83 & 0.5 & $2\cdot10^{-9}$ & $1.2\cdot10^{-8}$ & $7\cdot10^{-9}$ \\

\hline

 10 & -0.73 & 0.8& $3.2\cdot10^{-6}$ & $2.5\cdot10^{-5}$ & $2.7\cdot10^{-5}$ \\
 12 & -1 & 0.8 & $5.5\cdot10^{-8}$ & $2.7\cdot10^{-7}$ & $2.7\cdot10^{-7}$ \\
 14 & -1.31 & 0.8 & $4\cdot10^{-10}$ & $2\cdot10^{-9}$ & $2\cdot10^{-9}$ \\

\hline
\end{tabular}
\end{center}
\caption{Acceptance rates $r$ of R- and NR- algorithms for several spectra and several values of $E_{\hbox{\scriptsize av}}$ as described in the text. 
}
\end{table*}

As one can see from Table~\ref{tab:performance}, the value of acceptance rates $r$ for all algorithms decreases rapidly with the increase of $N$ to the extent that the sampling of Hilbert spaces with dimensions larger than 14 appears impractical with the present algorithms.  

\begin{figure}[t]
\begin{pspicture}(0,0)(5,5)
\put(0.5,0){ \epsfxsize= 2.7in \epsfbox{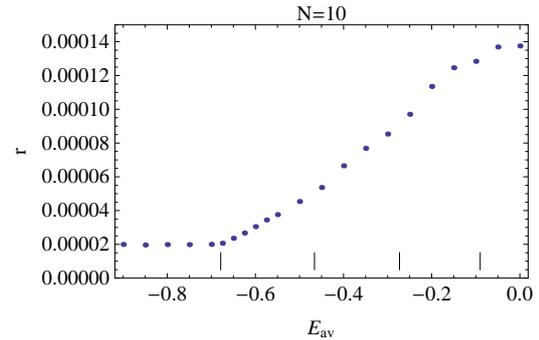} }
\end{pspicture}
\caption{Plot of the acceptance rate $r$ of the R-algorithm versus the average energy $E_{\hbox{\scriptsize av}}$ for the 10-level energy spectrum (\ref{spectrum10}).
}
\label{fig:rofeav}
\end{figure}

For a given spectrum, the acceptance rates also decrease with the increase of the ratio $E_{\hbox{\scriptsize av}}/E_1$. This fact is further illustrated in Fig.~\ref{fig:rofeav} for the R-algorithm applied to the spectrum with $N=10$. An interesting detail revealed by Fig.~\ref{fig:rofeav} is that the acceptance rate does not depend on $E_{\hbox{\scriptsize av}}$ for $E_1 < E_{\hbox{\scriptsize av}} < E_2$.

Comparing the relative performance of the three algorithms, it can be observed from Table~\ref{tab:performance}, that the optimisation of R- or NR- algorithm does not result in significant increase of the acceptance rates for the values of $E_{\hbox{\scriptsize av}}$ close to the centre of the spectrum. However, as the average energy deviates from the centre the optimisation leads to the increase of the acceptance rates by about a factor of 5. The acceptance rates for optimised R- and NR-algorithms are close to each other for all combinations of parameters considered.

The decrease of the acceptance rates with the departure of $E_{\hbox{\scriptsize av}}$ from the centre of the spectrum, as well as the insensitivity of the algorithms to the optimisation for $E_{\hbox{\scriptsize av}}$ close to zero, can probably be attributed to the fact that, according to section \ref{ch:vertices}, the number of vertices of manifold $M$ is maximal --- equal to $N^2/4$ --- for $E_{\hbox{\scriptsize av}} = 0$, and then it quickly decreases towards $N-1$ as  $E_{\hbox{\scriptsize av}}$ deviates from zero. Larger number of vertices, supposedly indicates that manifold $M$ has more even shape, which fills a larger volume fraction of any box around it irrespective of the box orientation. On the contrary, the smaller number of vertices may imply more uneven shape of manifold $M$ occupying a smaller fraction of any reasonably-shaped box, with the occupied fraction being strongly dependent on the box orientation.

\section{Conclusions}
\label{ch:summary}

This paper described and optimised two algorithms for performing Monte Carlo sampling of quantum superpositions in high-dimensional Hilbert spaces under the constraint on the energy expectation value. The two algorithms are distinguished by the shape of Monte-Carlo sampling boxes: either rectangular or parallelogram-like. The optimisation included finding the box orientations allowing smaller box sizes and, therefore, larger Monte-Carlo acceptance rates.  Both algorithms were found to exhibit similar performances in the optimised form.  The benefit of the optimisation in terms of algorithm's acceptance rate was found to be small for the values of  $E_{\hbox{\scriptsize av}}$ close to the centre of the Gaussian-like spectra considered  but increased by about factor of 5 as $E_{\hbox{\scriptsize av}}$ moved to the wings of the spectra. We have found that the largest dimension of Hilbert space that can realistically be explored with these algorithms is 14.

Our efforts to optimise the choice of the sampling boxes also indicate that for $E_{\hbox{\scriptsize av}}$ close to the centres of energy spectra, the manifolds of interest are more even-shaped, while, in the more interesting regime of $E_{\hbox{\scriptsize av}}$ being on the wings of the energy spectra, the resulting manifolds are more uneven and difficult to sample.

\appendix

\section{Volume of high-dimensional parallelogram-like box}
\label{ch:directcalc}

\begin{figure}[t]
\begin{pspicture}(-1.3,0)(6,2.5)
\pspolygon[linewidth=1pt](0,0.35)(2,2.35)(6,2.35)(4,0.35)
\rput(3,0.1){$b _2$}
\rput(0.3,1.6){$b _1$}

\psline[linewidth=1.5pt,linecolor=green](2,0.35)(2,2.35)
\rput[color=green](2.2,1.35){$h_1$}

\psline[linewidth=1.2pt]{->}(3.1,0.35)(3.1,1.35)
\rput(3.3,1.4){${\mathbf s}_1$}

\end{pspicture}
\caption{Cartoon of a two-dimensional parallelogram illustrating the calculation of the volume of high-dimensional parallelogram-like box. The volume (area) of this parallelogram is   $b _2  h_1$, where $h_1$ is the height and $b _2$ is the base side as indicated. }
\label{fig:parallelogram}
\end{figure}
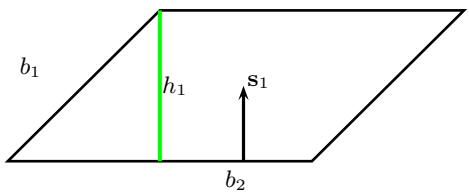

In order to determine the volume of an $(N-2)$-dimensional parallelogram-like box, the following algorithm can be applied iteratively.  Each iterative step of this procedure is illustrated by Fig.~\ref{fig:parallelogram}. 

The numerical routine begins by selecting one vertex as the origin and then determining all edges connecting the origin vertex to the $N-2$ neighbouring vertices. The vector basis which is defined by the set of all edges is denoted as $\{{\mathbf b}_1, {\mathbf b}_2\,\ldots {\mathbf b}_{N-2}\}$.
 
The first iteration takes the subset of the basis vectors $\{{\mathbf b}_{2}\,\ldots {\mathbf b}_{N-2}\}$ and finds a normalised vector ${\mathbf s}_1$ which is orthogonal to each vector of the above subset. This is done by solving a system of linear equations for the orthogonalisation conditions. Once ${\mathbf s}_1$ is obtained, the following height parameter can be defined:
\begin{equation}
h_1=({\mathbf b}_1 \cdot {\mathbf s}_1)
\end{equation}

The second iteration finds a normalised vector ${\mathbf s}_2$ which is orthogonal to $\{{\mathbf b}_{3}\,\ldots {\mathbf b}_{N-2}, {\mathbf s}_1\}$ and then finds the corresponding $h_2$. 

The iterations are continued until $h_{N-3}$ is found. The resulting volume of the box is then given by
\begin{equation}
\mathrm{V_B}=|{\mathbf b}_{N-2}| \cdot \prod_{i=1}^{N-3}{h_i}
\end{equation}

\section{Maximum volume box in the NR-algorithm}

\label{ch:maxvolbox}

According to the results of Section~\ref{ch:2ndalg}, the maximum sampling box volume for the NR-algorithm corresponds to the choice of the origin vertex $v_{i,j}$ such that $E_i$ and $E_j$ bracket the average energy $E_{\hbox{\scriptsize av}}$. This choice is special because of the geometrical factors described below .

The vertices connected by the common edges to $v_{i,j}$ form two groups: the first group consists of vertices $v_{i,m}$ (the same first index as for the origin vertex), while the second group consists of $v_{n,j}$ (the same second index as for the origin vertex). 
As discussed in Section~\ref{ch:vertices}, each vertex $v_{i,m}$ in the  first group has only two non-zero coordinates $p_i$ and $p_m$,
where:
\begin{eqnarray}
\label{eqn:pmpn}
p_i &= \frac{E_{\hbox{\scriptsize av}}-E_m}{E_i-E_m} \\
p_m &= \frac{E_{\hbox{\scriptsize av}}-E_i}{E_m -E_i}.
\end{eqnarray}
In this group, the typical situation is that $E_i$ is much closer to $E_{\hbox{\scriptsize av}}$ than $E_m$. Therefore, $p_i$ is close to 1, and $p_m$ is close to zero. In other words, most vertices of the first group cluster around the point $(0,...,0,1_i, 0, ...,0)$, where the notation $1_i$ indicates that $p_i=1$. Likewise the vertices of the second group cluster around $(0,...,0,1_j, 0, ...,0)$.
This clustering results in relatively small angles between all basis vectors connecting the origin vertex to the vertices of the same group, which presumably leads to 
a non-optimal choice of box $B$ around $M$. 

\bibliographystyle{unsrt}


\end{document}